# Multiscale Hilbert expansion and hydrodynamics beyond Navier-Stokes

O. M. Chekmareva* and I. B. Chekmarev**

A truncated Hilbert expansion with multiple scales is used to construct asymptotic solution of the linearized one-dimensional Boltzmann equation with small Knudsen number. The freedom ganed by introducing new "independent" time variables is exploited to eliminate the sources of secular behavior as the time interval increases. The regular gas-dynamic equations then appear as necessary conditions for expansion to be uniformly valid on a long time intervals.

## Introduction

The integro-differential Boltzmann equation (BE) is the base of the kinetic theory of dilute gas consisting of interacting particles without internal degrees of freedom [1-4]. The equation describes the time evolution of the one-particle distribution function providing all the information necessary to determine gas density, velocity, temperature, stresses, and heat flux. The BE is suitable at any Knudsen number Kn, which is the ratio between mean free path of a particle to a typical macroscopic length of the problem studied, and in the limit of small Kn also covers the area of classical gas dynamics. Therefore, solving the Boltzmann equation for small Kn and asymptotic transition from microscopic description to macroscopic with hydrodynamic quantities is one of the main problems of kinetic theory. For this reason, one of the tasks of the theory based on the Boltzmann equation is to develop some approximate macroscopic models and to determine the limits of their applicability. This long standing task is the subject of many extensive studies, started with the fundamental papers of Hilbert [1-2] and continuing further after the works of Chapman and Enskog [3]. As is well known, the literature dealing with the Boltzmann equation is immense. Our list includes mainly the works actually used by the writing this paper. Detailed reviews of various methods treated the problem, the results , advantages and disadvantages one can find in many books on the kinetic theory of gases, in particular, [3-5], as well as in [6-8] containing additional references and reviews.

In this paper, the truncated Hilbert expansion with multiscales is used to construct an asymptotic solution of the linearized one-dimensional Boltzmann equation in the hydrodynamic
limit. This problem, due to its relative simplicity, is well studied (in particular [6, 8, 23, 24]) and allows to test and compare various approximate methods for solving the Boltzmann equation.

The presence of a small parameter Kn in the BE enables the use of perturbation methods. The solution is often supposed to be a truncated series in powers of the small parameter with unknown coefficients depending on spatial and temporal variables. Since the error committed by retaining a few terms of the asymptotic expansion is of the same order as the first neglected term, the truncation assumes that this term is small compared to the previous ones. Whether this presumption will be fulfilled cannot be said in advance, since the contribution of each term of the expansion depends on its coefficient. Therefore, the principle question of how reasonable truncation is and when the solution can be applied is reduced to the analysis of the following higher-order terms. As it is known, direct power series fulfill this condition only on finite intervals and are not uniformly valid, since the higher-order terms grow with time and break our initial assumptions.

Improving the accuracy of approximation on a finite interval and finding a leading-order term uniformly valid on a large time interval are different problems. In the latter case, the zeroth-approximation must be regulated in such a way that it does not cause secularities in subsequent approximations. The method of multiple scales is well adapted to get around these difficulties. The principle point of the method is to identify the sources of secular behavior and, utilizing the freedom gained by introducing new time scales, to keep the expansion uniformly valid as the time interval is extended. The general theory and numerous examples of its successful application to various problems can be found in [9–12] and, in particular, in [13–16].

The method of multiple scales can also be applied to the Boltzmann equation and its moments. To demonstrate this approach in detail, we consider a time evolution of small disturbances in a monatomic gas with the goal to obtain the lowest-order term of expansion that will be uniformly valid as the time interval is extended. Such truncation will be justified only if the next (first) term of the asymptotic expansion remains small compared to this leading one over the entire time range of interest. For this purpose, we introduce a set of linear time scales which will be determined by the requirments that no secular terms arise in subsequent perturbations as the time interval is extended. The regular equations of gas-dynamics appear then as a consequence of uniformizing expansion in a small Kn.

In particular, it is shown that the application of multiple-time scales technique to the linearized Boltzmann equation leads to the equations governing the hydrodynamic quantities beyond the Navier–Stokes level with uniform error $O(\varepsilon)$ over the time interval $O(\varepsilon) < t \leqslant \varepsilon^{-2}$. One equation corresponds to the heat conduction mode and the other two linearized Burgers–KdV equations describe two separate sound waves propagating either in the positive or negative $x$-direction. The dispersion relations following from them are identical to those found by L. Sirovich in the hydrodynamic limit using 13 moments equations obtained by expanding the perturbed distribution functions of the linearized Boltzmann equation in the eigenfunctions of the collision operator [25].

## The problem and asymptotic procedure

It is supposed that a simple gas at rest with the constant density $n_{s0}$ and the constant temperature $T_s$ occupies the hole physical space. We take as scales for the distribution function $f$, particle velocity $c$, time $t$, radius vector $r$ and collision parameter $b$ the quantities.

$$f_s = \frac{n_{s0}}{C_s^3}, \quad C_s = \sqrt{\frac{kT_s}{m}}, \quad t_s, \quad L_s, \quad a,$$

where $m$ is the particle mass, $t_s$, is a typical macroscopic time, $L_s$ is a typical macroscopic size, $a$ is the effective particle radius [18]. Let us note that this scales collection is convenient to study weak nonequilibrium processes. Keeping the same letter designations for dimensional and dimensionless quantities, we transform one-dimensional Boltzmann equation for monatonic gas to dimensionless form

$$\varepsilon(Sh\frac{\partial f}{\partial t} + c_x\frac{\partial f}{\partial x}) = Q(f,f) = \int [f(c'_1)f(c') - f(c_1)f(c)]gbdbd\beta d c_1.$$

Here, we have used the conventional notation $\varepsilon = l_s/L_s$, $l_s = 1/n_{s0}\pi a^2$ and assumed Knudsen number $\varepsilon$ to be small, i.e., $\varepsilon \ll 1$. The parameter $Sh = L_s/C_s t_s$, the so-called Strouhal number. is the ratio of the time it takes for a sound wave to travel a distance $L_s$ to the time scale $t_s$. The $Sh$ number largely determines the type of damping of disturbances in a gas. For small values of the Strouhal number (quasi-stationary regime), the damping has a monotonous character and is described by the incompressible Navier-Stokes equations. Of a special interest, which will be considered in the article, is the case of $Sh=O(1)$, when the gas behaves as an elastic medium. Thus,

$$\varepsilon(\frac{\partial f}{\partial t}+c_x\frac{\partial f}{\partial x})=Q(f,f).$$

To start, an initial disturbance with a scale density $n_{s1}$ is introduced, which supposed such small relative to $n_{s0}$ that $\delta=n_{s1}/n_{s0}\ll\varepsilon$. We confine ourselves to the the linear approximation

$$f=f_M(1+\delta\varphi)$$

and consider the linearized one-dimensional Boltzmann equation

$$\varepsilon(\frac{\partial\varphi}{\partial t}+c_x\frac{\partial\varphi}{\partial x})=L\varphi=\int f_M(\boldsymbol{c_1})[\varphi]gbdbd\beta d\boldsymbol{c_1},\quad \varepsilon\ll 1, \tag{01}$$

$$[\varphi]=\varphi(\boldsymbol{c'}_1)+\varphi(\boldsymbol{c'})-\varphi(\boldsymbol{c}_1)-\varphi(\boldsymbol{c}),$$

while the undisturbed gas is supposed to be at rest and has a global Maxwellian

$$f_M=\frac{1}{(2\pi)^{3/2}}\exp(-\frac{c^2}{2}). \tag{0.2}$$

In terms of the distribution function the macroscopic quantities are defined as

$$\int f_M\varphi d\boldsymbol{c}=n,\quad \int c_x f_M\varphi d\boldsymbol{c}=u,\quad \int \frac{c^2}{2}f_M\varphi d\boldsymbol{c}=\frac{3}{2}p. \tag{0.3}$$

The linearized Boltzmann collision operator [4]

$$L\varphi=\int f_M(\boldsymbol{c_1})[\varphi(\boldsymbol{c'}_1)+\varphi(\boldsymbol{c'})-\varphi(\boldsymbol{c}_1)-\varphi(\boldsymbol{c})]gbdbd\beta d\boldsymbol{c_1}$$

is self-adjoint, i.e.,

$$(\chi,L\varphi)=(L\chi,\varphi),$$
$$(\chi,L\varphi)=-\int d\boldsymbol{c}\int f_0(\boldsymbol{c})f_0(\boldsymbol{c}_1)[\chi][\varphi]gbdbd\beta d\boldsymbol{c}_1, \tag{04}$$
$$[\chi]=[\chi'_1+\chi'-\chi_1-\chi],\quad [\varphi]=[\varphi'_1+\varphi'-\varphi_1-\varphi],$$

has eigenfunctions $\psi_{rl}$ and eigenvalues $\lambda_{rl}$ defined as

$$L\psi_{rl}=\lambda_{rl}\psi_{rl}. \tag{0.5}$$

The first five eigenvalues corresponding to eigenfunctions $1$, $\boldsymbol{c}$, $c^2$ are equal to zero. Hence, the solution of the homogeneous equation is a linear combination of these five eigenfunctions

$$Lg=0, \quad g=\alpha+\beta_i c_i+\gamma \frac{c^2}{2}. \tag{06}$$

The next five eigenvalues are negative and for Maxwell molecules are [24]:

$$\lambda_{11}=\frac{2}{3}\lambda_{02}, \quad \lambda_{03}=\frac{3}{2}\lambda_{02}, \quad \lambda_{20}=\frac{2}{3}\lambda_{02}, \quad \lambda_{12}=\frac{7}{6}\lambda_{02}, \tag{0.7}$$

and correspond to the five integral equations

$$L\psi_{02}=\lambda_{02}\psi_{02}, \quad L\psi_{11}=\frac{2}{3}\lambda_{02}\psi_{11}, \quad L\psi_{03}=\frac{3}{2}\lambda_{02}\psi_{03},$$

$$L\psi_{20}=\frac{2}{3}\lambda_{02}\psi_{20}, \quad L\psi_{12}=\frac{7}{6}\lambda_{02}\psi_{12}, \tag{0.8}$$

where

$$\psi_{01}=c^2-\frac{3}{2}, \quad \psi_{02}=(c_x^2-\frac{c^2}{3}), \quad \psi_{11}=c_x(\frac{c^2}{2}-\frac{5}{2}), \quad \psi_{03}=c_x(c_x^2-\frac{3}{5}c^2),$$

$$\psi_{20}=(\frac{15}{4}-\frac{5}{2}c^2+\frac{c^4}{4}), \quad \psi_{12}=(c_x^2-\frac{c^2}{3})(\frac{c^2}{2}-\frac{7}{2}). \tag{09}$$

These eigenfunctions are orthogonal, i.e., $\int f_0\psi_{kl}\psi_{k'l'}d\boldsymbol{c}=0$, $k\neq k'$, $l\neq l'$ with the following normalization factors:

$$\int f_0\psi_{02}^2 d\boldsymbol{c}=\frac{4}{3}, \quad \int f_0\psi_{11}^2 d\boldsymbol{c}=\frac{5}{2},$$

$$\int f_0\psi_{12}^2 d\boldsymbol{c}=\frac{14}{3}, \quad \int f_0\psi_{20}^2 d\boldsymbol{c}=\frac{15}{2}, \quad \int f_0\psi_{03}^2 d\boldsymbol{c}=\frac{12}{5} \tag{0.10}$$

and satisfy the recursion relations:

$$c_x\psi_{02}=\psi_{03}+\frac{8}{15}\psi_{11}+\frac{4}{3}c_x,$$

$$c_x\psi_{11}=\psi_{12}+\psi_{02}+\frac{2}{3}\psi_{20}+\frac{5}{3}(\frac{c^2}{2}-\frac{3}{2}). \tag{0.11}$$

There are various forms of the multiscale method [9-12]. We introduce, as in [16-20], a sequence of time scales $t_k=\varepsilon^k t$ and assume that the solution to Eq. (0.1) can be found in the form

$$\varphi = \sum_{k=0}^{N-1} \varepsilon^k \varphi_k(x, t_0, t_1, t_2, ..t_k) + O(\varepsilon^N) \qquad (0.12)$$

that implies the related expansions

$$n = n_0 + \varepsilon n_1 + \varepsilon^2 n_2 + ..., \quad u = u_0 + \varepsilon u_1 + \varepsilon^2 u_2 + ..., \quad p = p_0 + \varepsilon p_1 + \varepsilon^2 p_2 + ..., \qquad (0.13)$$

$$\int f_0 \varphi_k \, d\mathbf{c} = n_k, \quad \int c_x f_0 \varphi_k \, d\mathbf{c} = u_k, \quad \int \frac{c^2}{2} f_0 \varphi_k \, d\mathbf{c} = \frac{3}{2} p_k. \qquad (0.14)$$

According to the terminology used, the expansion (0.12) is outer approximation and describes weak nonequilibrium processes. High nonequilibrium kinetic initial and boundary domains are not considered in our paper.

Since $t_k$ are treated as new independent variables, the time derivative is transformed as follows:

$$\frac{\partial}{\partial t} = \frac{\partial}{\partial t_0} + \varepsilon \frac{\partial}{\partial t_1} + \varepsilon^2 \frac{\partial}{\partial t_2} + \varepsilon^3 \frac{\partial}{\partial t_3} + ... \, . \qquad (0.15)$$

Substituting (0.12) and (0.15) into (0.1) and equating the coefficient of each power of $\varepsilon$ to zero yield the basic hierarchy of linear integral equations which determine successively $\varphi_k$

$$L \varphi_k = Q_k, \qquad (0.16)$$

where $Q_k$ are also constructed as in Hilbert expansion by means of the previous approximations:

$$\begin{aligned}
O(\varepsilon^0): &\quad Q_0 = 0, \\
O(\varepsilon^1): &\quad Q_1 = \frac{\partial \varphi_0}{\partial t_0} + c_x \frac{\partial \varphi_0}{\partial x}, \\
O(\varepsilon^2): &\quad Q_2 = \frac{\partial \varphi_0}{\partial t_1} + \frac{\partial \varphi_1}{\partial t_0} + c_x \frac{\partial \varphi_1}{\partial x}, \\
O(\varepsilon^3): &\quad Q_3 = \frac{\partial \varphi_0}{\partial t_2} + \frac{\partial \varphi_1}{\partial t_1} + \frac{\partial \varphi_2}{\partial t_0} + c_x \frac{\partial \varphi_2}{\partial x} ...
\end{aligned} \qquad (0.17)$$

However, the set (0.16) differs essentially from the Hilbert expansion and involves arbitrary functions of the times $t_k$. To determine their dependence on these new variables, we use the freedom gained by (0.15) to impose additional constraints. Our only explicit restrictions will be: if the expansions (0.12-0.13) are to be valid for a long time, i.e., the asymptotic solution of the problem may be represented by the first terms of the expansion, it is necessary to require that each term of the expansion be no more singular than the preceding one for all times of interest. This condition is equivalent to the elimination of secular terms.

The solutions of Eqs. (0.16) can be represented in the form

$$\varphi_k = \alpha_k + \beta_{ki} c_i + \gamma_k \frac{c^2}{2} + h_k, \qquad (0.18)$$

where $h_k$ is a particular solution of the nonhomogeneous equation (0.16).

At each step arbitrary functions $\alpha_k, \beta_{ki}, \gamma_k$ are defined by the solvability conditions:

$$\int f_M Q_k d\mathbf{c}=0, \quad \int f_M c_x Q_k d\mathbf{c}=0, \quad \int f_M c^2 Q_k d\mathbf{c}=0. \tag{0.19}$$

As it follows from (0.12-0.14), functions $\alpha_k, \beta_k, \gamma_k$ may be expressed also in terms of the macroscopic quantities

$$n_k=\alpha_k+\frac{3}{2}\gamma_k, \quad \frac{5}{2}p_k=\frac{3}{2}(\alpha_k+\frac{5}{2}\gamma_k),$$

$$-\alpha_k=\frac{3}{2}p_k-\frac{5}{2}n_k=s_k, \quad \gamma_k=T_k, \quad \beta_k=u_k.$$

Accordingly, we can write $\varphi_k$ as

$$\varphi_k=g_k+h_k=-s_k+u_k c_x+T_k\frac{c^2}{2}+h_k, \quad T_k=p_k-n_k, \quad s_k=\frac{3}{2}p_k-\frac{5}{2}n_k. \tag{0.20}$$

Since at the order $\varepsilon^0$ the integral equation $L\varphi_0=0$, its solution is the linear combination of the invariants given (0.20)

$$\varphi_0=-s_0+u_0 c_x+T_0\frac{c^2}{2}, \tag{0.21}$$

where $s_0, u_0, T_0$ are yet unknown functions of $x$ and $t_k$.

The main our task will be to determine uniform asymptotic solution of equation (0.1) with error $O(\varepsilon)$ for the times $O(\varepsilon)<t\leqslant\varepsilon^{-2}$. Such truncation will be justified only if $\varphi_1$ remains limited independently of time.

### The first approximation

The solvability conditions to the first order equation

$$\frac{\partial\varphi_0}{\partial t_0}+c_x\frac{\partial\varphi_0}{\partial x}=L\varphi_1 \tag{1.1}$$

leads to the linearized Euler's equations with respect to the fast time $t_0$

$$\frac{\partial n_0}{\partial t_0}+\frac{\partial u_0}{\partial x}=0, \quad \frac{\partial u_0}{\partial t_0}+\frac{\partial p_0}{\partial x}=0, \quad \frac{\partial p_0}{\partial t_0}+\frac{5}{3}\frac{\partial u_0}{\partial x}=0, \tag{1.2}$$

which reduce to the wave equations for $u_0$ and $p_0$

$$\frac{\partial^2 u_0}{\partial t_0^2} - a_0^2 \frac{\partial^2 u_0}{\partial x^2} = \frac{\partial^2 p_0}{\partial t_0^2} - a_0^2 \frac{\partial^2 p_0}{\partial x^2} = 0, \quad a_0^2 = \frac{5}{3} \tag{1.3}$$

and the adiabatic relation

$$\frac{\partial s_0}{\partial t_0} = 0, \quad s_0 = \frac{3}{2} p_0 - \frac{5}{2} n_0. \tag{1.4}$$

The general solution of the integral equation (1.1) may be written as

$$\varphi_1 = -s_1 + u_1 c_x + \frac{c^2}{2} T_1 + h_1, \quad s_1 = \frac{3}{2} p_1 - \frac{5}{2} n_1, \quad \frac{5}{2} T_1 = p_1 + s_1, \tag{1.5}$$

where $h_1$ is a particular solution of (1.1).

Substituting (0.20) into (1.1) and using (1.2) gives the integral equation for $h_1$

$$L h_1 = \frac{\partial T_0}{\partial x} \psi_{11} + \frac{\partial u_0}{\partial x} \psi_{02}, \quad \psi_{02} = \left(c_x^2 - \frac{c^2}{3}\right), \quad \psi_{11} = c_x \left(\frac{c^2}{2} - \frac{5}{2}\right),$$

which because of Eqs. (0.8), may be rewrite as

$$\frac{1}{\lambda_{11}} \frac{\partial T_0}{\partial x} L \psi_{11} + \frac{1}{\lambda_{02}} \frac{\partial u_0}{\partial x} L \psi_{02} = L h_1.$$

Thus,

$$h_1 = \frac{1}{\lambda_{11}} \frac{\partial T_0}{\partial x} \psi_{11} + \frac{1}{\lambda_{02}} \frac{\partial u_0}{\partial x} \psi_{02}. \tag{1.6}$$

Note, $\varphi_0$ is as yet unknown function of the slow times $t_k$ to be determined by the error estimate in the next orders.

**The second approximation**

We go on to the next order and consider the integral equation

$$\frac{\partial \varphi_0}{\partial t_1} + \frac{\partial \varphi_1}{\partial t_0} + c_x \frac{\partial \varphi_1}{\partial x} = L \varphi_2 \tag{2.1}$$

with solvability conditions

$$\frac{\partial u_0}{\partial t_1} + \frac{\partial u_1}{\partial t_0} + \frac{\partial p_1}{\partial x} = -\frac{\partial}{\partial x} \int f_0 h_1 \psi_{02} d\,c, \tag{2.2a}$$

$$\frac{\partial p_0}{\partial t_1} + \frac{\partial p_1}{\partial t_0} + \frac{5}{3} \frac{\partial u_1}{\partial x} = -\frac{2}{3} \frac{\partial}{\partial x} \int f_0 h_1 \psi_{11} d\,c, \tag{2.2b}$$

$$\frac{\partial s_0}{\partial t_1}+\frac{\partial s_1}{\partial t_0}+\frac{\partial}{\partial x}\int f_0 h_1 \psi_{11} d\,c. \qquad (2.2c)$$

Note that the compatibility conditions thus obtained are not ordinary conservation equations, but equations in which the expansion is based on a scaling given by (0.15).

Substitution (1.6) into (2.2c) yields

$$-\frac{\partial}{\partial x}\int f_0 h_1 \psi_{02} d\,c = -\frac{4}{3\lambda_{02}}\frac{\partial^2 u_0}{\partial x^2}, \qquad (2.3a)$$

$$-\frac{2}{3}\frac{\partial}{\partial x}\int f_0 h_1 \psi_{11} d\,c = -\frac{5}{3\lambda_{11}}\frac{\partial^2 T_0}{\partial x^2}, \qquad (2.3b)$$

$$\frac{\partial s_0}{\partial t_1}+\frac{\partial s_1}{\partial t_0} = -\frac{5}{2\lambda_{11}}\frac{\partial^2 T_0}{\partial x^2} = -\frac{1}{\lambda_{11}}\frac{\partial^2 (p_0+s_0)}{\partial x^2}=0. \qquad (2.2c)$$

Since $s_0$ doesn't depend on $t_0$, it follows then from (2.2c) that

$$\frac{\partial s_0}{\partial t_1}+\frac{1}{\lambda_{11}}\frac{\partial^2 s_0}{\partial x^2}=0, \quad \frac{\partial}{\partial t_0}(s_1-\frac{1}{\lambda_{11}}\frac{\partial u_0}{\partial x})=0. \qquad (2.4)$$

The Eqs. (2.2a-b) accordingly become

$$\frac{\partial u_0}{\partial t_1}+\frac{\partial \tilde{u}_1}{\partial t_0}+\frac{\partial p_1}{\partial x}=-\frac{4}{3\lambda_{02}}\frac{\partial^2 u_0}{\partial x^2},$$

$$\frac{\partial p_0}{\partial t_1}+\frac{\partial p_1}{\partial t_0}+\frac{5}{3}\frac{\partial \tilde{u}_1}{\partial x}=-\frac{2}{3\lambda_{11}}\frac{\partial^2 p_0}{\partial x^2}, \quad \tilde{u}_1=u_1+\frac{2}{5\lambda_{11}}\frac{\partial s_0}{\partial x}, \quad \frac{\partial s_0}{\partial t_0}=0$$

and lead to the nonhomogeneous equations for $\tilde{u}_1$ and $p_1$

$$\frac{\partial^2 \tilde{u}_1}{\partial t_0^2}-\frac{5}{3}\frac{\partial^2 \tilde{u}_1}{\partial x^2}=-\frac{\partial}{\partial t_0}[2\frac{\partial u_0}{\partial t_1}+(\frac{4}{3\lambda_{02}}+\frac{2}{3\lambda_{11}})\frac{\partial^2 u_0}{\partial x^2}], \qquad (2.5a)$$

$$\frac{\partial^2 p_1}{\partial t_0^2}-\frac{5}{3}\frac{\partial^2 p_1}{\partial x^2}=-\frac{\partial}{\partial t_0}[2\frac{\partial p_0}{\partial t_1}+(\frac{4}{3\lambda_{02}}+\frac{2}{3\lambda_{11}})\frac{\partial^2 p_0}{\partial x^2}]. \qquad (2.5b)$$

The terms on the right-hand sides of (2.5) produce particular solutions proportional to $t_0$ and make $\varepsilon\varphi_1$ of the same order as $\varphi_0$ for $t=O(\varepsilon^{-1})$, unless to require

$$\frac{\partial u_0}{\partial t_1}+(\frac{2}{3\lambda_{02}}+\frac{1}{3\lambda_{11}})\frac{\partial^2 u_0}{\partial x^2}=0\ ,\quad \frac{\partial p_0}{\partial t_1}+(\frac{2}{3\lambda_{02}}+\frac{1}{3\lambda_{11}})\frac{\partial^2 p_0}{\partial x^2}=0\ . \qquad (2,6)$$

Thus, Eqs. (2.6) determine the behavior of $\varphi_0$ with respect to the time $t_1$.

Expressing the time derivatives $\partial/\partial t_0$ and $\partial/\partial t_1$ in (0.15) in terms of space derivatives according to (1.2) and (2.6), we obtain the following equations governing $u$ and $p$ with error of $O(\varepsilon)$ valid for $t \leqslant \varepsilon^{-1}$:

$$\frac{\partial u}{\partial t}+\frac{\partial p}{\partial x}+\varepsilon(\frac{2}{3\lambda_{02}}+\frac{1}{3\lambda_{11}})\frac{\partial^2 u}{\partial x^2}=0\ ,$$

$$\frac{\partial p}{\partial t}+\frac{5}{3}\frac{\partial u}{\partial x}+\varepsilon(\frac{2}{3\lambda_{02}}+\frac{1}{3\lambda_{11}})\frac{\partial^2 p}{\partial x^2}=0\ ,\quad \frac{\partial s}{\partial t}+\frac{1}{\lambda_{11}}\varepsilon\frac{\partial^2 s}{\partial x^2}=0\ , \qquad (2.7)$$

$$u=u_0+O(\varepsilon),\ p=p_0+O(\varepsilon),\ s=s_0+O(\varepsilon),\ t\leqslant \varepsilon^{-1}.$$

It should be remarked here that the error in approximations of (2.7) will not be of $O(\varepsilon)$ if the time interval is longer than $O(\varepsilon^{-1})$.

The Eqs. (2.7) differ from the linearized Navier-Stokes system. In the Navier–Stokes theory the viscosity and heat conduction terms are included on the basis of the experimental Hook-Newton and Fourier laws. As it is known, the transport coefficients in gas are turn out to be proportional to the free path [26]. As a result, the nondimensional Navier-Stokes equations in the case of gas contain themselves the small parameter and can lead to singular solutions. As it can easily be proved, the two-time scale method applied to the singular Navier-Stokes set in the limiting case of small $\varepsilon$ gives the asymptotic result in full agreement with the regular set (2.7). Thus, within the framework of the models used, the Boltzmann equation and the Navier-Stokes system demonstrate only asymptotic equivalence.

After eliminating secular terms, Eqs. (2.2a,b) reduce to the linearized nonhomogeneous Euler's equations for the next approximation with respect to the fast time $t_0$

$$\frac{\partial \tilde{u}_1}{\partial t_0}+\frac{\partial p_1}{\partial x}=-(\frac{2}{3\lambda_{02}}-\frac{1}{3\lambda_{11}})\frac{\partial^2 u_0}{\partial x^2}\ ,$$

$$\frac{\partial p_1}{\partial t_0}+\frac{5}{3}\frac{\partial \tilde{u}_1}{\partial x}=(\frac{2}{3\lambda_{02}}-\frac{1}{3\lambda_{11}})\frac{\partial^2 p_0}{\partial x^2}\ . \qquad (2.8)$$

### The third approximation

To extend the validity of $\varphi_0$ up to times $O(\varepsilon^{-2})$, we turn to the third-order equation

$$\frac{\partial \varphi_0}{\partial t_2}+\frac{\partial \varphi_1}{\partial t_1}+\frac{\partial \varphi_2}{\partial t_0}+c_x\frac{\partial \varphi_2}{\partial x}=L\varphi_3 \qquad (3.1)$$

and examine whether $\varphi_1$ will contain secular terms as time increases to $O(\varepsilon^{-2})$.

Since the collision operator $L$ is self-adjoint, we write compatibility conditions as

$$\frac{\partial u_0}{\partial t_2}+\frac{\partial u_1}{\partial t_1}+\frac{\partial u_2}{\partial t_0}+\frac{\partial p_2}{\partial x}=-\frac{\partial}{\partial x}\int f_0\psi_{02}\varphi_2 d\mathbf{c}=-\frac{1}{\lambda_{02}}\frac{\partial}{\partial x}\int f_0\psi_{02}L\varphi_2 d\mathbf{c},\qquad(3.2a)$$

$$\frac{\partial p_0}{\partial t_2}+\frac{\partial p_1}{\partial t_1}+\frac{\partial p_2}{\partial t_0}+\frac{5}{3}\frac{\partial u_2}{\partial x}=-\frac{2}{3}\frac{\partial}{\partial x}\int f_0\psi_{11}\varphi_2 d\mathbf{c}=-\frac{2}{3\lambda_{11}}\frac{\partial}{\partial x}\int f_0\psi_{11}L\varphi_2 d\mathbf{c},\qquad(3.2b)$$

$$\frac{\partial s_0}{\partial t_2}+\frac{\partial s_1}{\partial t_1}+\frac{\partial s_2}{\partial t_0}=-\frac{\partial}{\partial x}\int f_0 h_2\psi_{11}d\mathbf{c}=-\frac{1}{\lambda_{11}}\frac{\partial}{\partial x}\int f_0\psi_{11}L\varphi_2 d\mathbf{c}.\qquad(3.2c)$$

The above integals can be evaluated using the second-order distribution function obtained before and the recurrence relations (0.11)

$$\int f_0\psi_{02}l\varphi_2 d\mathbf{c}=\int f_0\psi_{02}(\frac{\partial\varphi_1}{\partial t_0}+c_x\frac{\partial\varphi_1}{\partial x})d\mathbf{c}=\int f_0[\psi_{02}\frac{\partial h_1}{\partial t_0}+(\frac{8}{15}\psi_{11}+\frac{4}{3}c_x)\frac{\partial\varphi_1}{\partial x}]d\mathbf{c},$$

$$\int f_0\psi_{11}L\varphi_2 d\mathbf{c}=\int f_0[\psi_{11}\frac{\partial h_1}{\partial t_0}+(\psi_{02}+\frac{5}{3}(\frac{c^2}{2}-\frac{3}{2}))\frac{\partial\varphi_1}{\partial x}]d\mathbf{c}.$$

It easy to verify that

$$\int f_0\psi_{02}L\varphi_2 d\mathbf{c}=\frac{4}{3}\frac{\partial\tilde{u}_1}{\partial x}-(\frac{4}{3\lambda_{02}}-\frac{8}{15\lambda_{11}})\frac{\partial^2 p_0}{\partial x^2},\qquad(3.3)$$

$$\int f_0\psi_{11}L\varphi_2 d\mathbf{c}=\frac{\partial p_1}{\partial x}+\frac{\partial}{\partial x}(s_1-\frac{1}{\lambda_{11}}\frac{\partial u_0}{\partial x})+(\frac{4}{3\lambda_{02}}-\frac{2}{3\lambda_{11}})\frac{\partial^2 u_0}{\partial x^2},\qquad(3.4)$$

where the integral (3.4) according to (2.8) can also be expressed in the form

$$\int f_0\psi_{11}L\varphi_2 d\mathbf{c}=-\frac{\partial u_1}{\partial t_0}-\frac{\partial u_0}{\partial t_1}+\frac{\partial}{\partial x}(s_1-\frac{1}{\lambda_{11}}\frac{\partial u_0}{\partial x})-\frac{2}{3\lambda_{11}}\frac{\partial^2 u_0}{\partial x^2}.\qquad(3.5)$$

Then Eq. (3.2c) may be written as follows:

$$\frac{\partial s_0}{\partial t_2}+(\frac{\partial}{\partial t_1}+\frac{1}{\lambda_{11}}\frac{\partial^2}{\partial x^2})(s_1-\frac{1}{\lambda_{11}}\frac{\partial u_0}{\partial x})+\frac{\partial}{\partial t_0}(s_2-\frac{1}{\lambda_{11}}\frac{\partial u_1}{\partial x}+\frac{1}{\lambda_{11}^2}\frac{\partial^2 T_0}{\partial x^2})=0.$$

Hence, because of (1.4) and (2.4), we can conclude that

$$\frac{\partial s_0}{\partial t_2}=0,\quad(\frac{\partial}{\partial t_1}+\frac{1}{\lambda_{11}}\frac{\partial^2}{\partial x^2})(s_1-\frac{1}{\lambda_{11}}\frac{\partial u_0}{\partial x})=0,\quad\frac{\partial}{\partial t_0}(s_2-\frac{1}{\lambda_{11}}\frac{\partial u_1}{\partial x}+\frac{1}{\lambda_{11}^2}\frac{\partial^2 T_0}{\partial x^2})=0.\qquad(3.6)$$

As a result, the two equations (3.2a-b) become

$$\frac{\partial u_0}{\partial t_2}+\frac{\partial u_1}{\partial t_1}+\frac{\partial \tilde{u}_2}{\partial t_0}+\frac{\partial p_2}{\partial x}=-\frac{1}{\lambda_{02}}\frac{\partial}{\partial x}\left[\frac{4}{3}\frac{\partial \tilde{u}_1}{\partial x}-\left(\frac{4}{3\lambda_{02}}-\frac{8}{15\lambda_{11}}\right)\frac{\partial^2 p_0}{\partial x^2}\right], \qquad (3.6a)$$

$$\frac{\partial p_0}{\partial t_2}+\frac{\partial p_1}{\partial t_1}+\frac{\partial p_2}{\partial t_0}+\frac{5}{3}\frac{\partial \tilde{u}_2}{\partial x}=-\frac{2}{3\lambda_{11}}\frac{\partial}{\partial x}\left[\frac{\partial p_1}{\partial x}+\left(\frac{4}{3\lambda_{02}}-\frac{2}{3\lambda_{11}}\right)\frac{\partial^2 u_0}{\partial x^2}\right], \qquad (3.6b)$$

where $\tilde{u}_1=u_1+\dfrac{2}{5\lambda_{11}}\dfrac{\partial s_0}{\partial x}$, $\tilde{u}_2=u_2+\dfrac{2}{5\lambda_{11}}\dfrac{\partial}{\partial x}\left(s_1-\dfrac{1}{\lambda_{11}}\dfrac{\partial u_0}{\partial x}\right)$.

As before, we obtain from (3.6) the nonhomogeneous wave equation for $\tilde{u}_2$

$$2\frac{\partial}{\partial t_0}\left(\frac{\partial u_0}{\partial t_2}+\frac{\partial u_1}{\partial t_1}\right)-\frac{\partial}{\partial t_1}\left(\frac{\partial \tilde{u}_1}{\partial t_0}+\frac{\partial p_1}{\partial x}\right)+\Box \tilde{u}_2=$$

$$=-\frac{\partial}{\partial t_0}\left[\left(\frac{4}{3\lambda_{02}}+\frac{2}{3\lambda_{11}}\right)\frac{\partial^2 \tilde{u}_1}{\partial x^2}-\left(\frac{4}{3\lambda_{02}^2}-\frac{8}{15\lambda_{02}\lambda_{11}}\right)\frac{\partial^3 p_0}{\partial x^3}\right]+$$

$$+\frac{2}{3\lambda_{11}}\frac{\partial^2}{\partial x^2}\left[\left(\frac{\partial \tilde{u}_1}{\partial t_0}+\frac{\partial p_1}{\partial x}\right)+\left(\frac{4}{3\lambda_{02}}-\frac{2}{3\lambda_{11}}\right)\frac{\partial^2 u_0}{\partial x^2}\right], \quad \Box=\frac{\partial^2}{\partial t_0^2}-a_0^2\frac{\partial^2}{\partial x^2}.$$

Using Eq. (2.8), as well as the constraints (2.6) and Eq. (1.2), we find finally after some algebra

$$\Box \tilde{u}_2=-2\frac{\partial}{\partial t_0}\left[\frac{\partial \tilde{u}_1}{\partial t_1}+\left(\frac{2}{3\lambda_{02}}+\frac{1}{3\lambda_{11}}\right)\frac{\partial^2 \tilde{u}_1}{\partial x^2}+\frac{\partial u_0}{\partial t_2}-\left(\frac{8}{15\lambda_{02}^2}-\frac{2}{5\lambda_{02}\lambda_{11}}+\frac{1}{10\lambda_{11}^2}\right)\frac{\partial^3 p_0}{\partial x^3}\right]. \qquad (3.7a)$$

Similarly, we obtain

$$\Box p_2=-2\frac{\partial}{\partial t_0}\left[\frac{\partial p_1}{\partial t_1}+\left(\frac{1}{3\lambda_{11}}+\frac{2}{3\lambda_{02}}\right)\frac{\partial^2 p_1}{\partial x^2}+\frac{\partial p_0}{\partial t_2}-\left(\frac{8}{9\lambda_{02}^2}-\frac{2}{3\lambda_{02}\lambda_{11}}+\frac{1}{6\lambda_{11}^2}\right)\frac{\partial^2 u_0}{\partial x^2}\right]. \qquad (3.7b)$$

The expression on the right-hand sides of (3.7) produces the secular solutions which grow proportional to $t_0$ and make $\varepsilon^2 u_2$ and $\varepsilon^2 p_2$ the same order as $\varepsilon u_1$ and $\varepsilon p_1$ when $t_0$ is as large as $O(\varepsilon^{-1})$. Hence, it is necessary to require

$$\frac{\partial \tilde{u}_1}{\partial t_1}+\left(\frac{2}{3\lambda_{02}}+\frac{1}{3\lambda_{11}}\right)\frac{\partial^2 \tilde{u}_1}{\partial x^2}=-\frac{\partial u_0}{\partial t_2}+\left(\frac{8}{15\lambda_{02}^2}-\frac{2}{5\lambda_{11}\lambda_{02}}+\frac{1}{10\lambda_{11}^2}\right)\frac{\partial^3 p_0}{\partial x^3}, \qquad (3.8a)$$

$$\frac{\partial p_1}{\partial t_1}+\left(\frac{2}{3\lambda_{02}}+\frac{1}{3\lambda_{11}}\right)\frac{\partial^2 p_1}{\partial x^2}=-\frac{\partial p_0}{\partial t_2}+\left(\frac{8}{9\lambda_{02}^2}-\frac{2}{3\lambda_{11}\lambda_{02}}+\frac{1}{6\lambda_{11}^2}\right)\frac{\partial^3 p_0}{\partial x^3}. \qquad (3.8b)$$

Thus, providing an error of order $\varepsilon^2$, we have improved the accuracy of the asymptotic solution to $O(\varepsilon)$ up to times of $O(\varepsilon^{-1})$. However, the right sides of (3.8) make $\tilde{u}_1$ and $p_1$ proportional to $t_1$. So if $t$ increases to $\varepsilon^{-2}$, the terms of $O(\varepsilon)$ are compared with the zeroth-term, unless

$$\frac{\partial u_0}{\partial t_2} - \left(\frac{8}{15\lambda_{02}^2} - \frac{2}{5\lambda_{02}\lambda_{11}} + \frac{1}{10\lambda_{11}^2}\right)\frac{\partial^3 p_0}{\partial x^3} = 0, \tag{3.9a}$$

$$\frac{\partial p_0}{\partial t_2} - \left(\frac{8}{9\lambda_{02}^2} - \frac{2\lambda_{02}}{3\lambda_{02}\lambda_{11}} + \frac{1}{6\lambda_{11}^2}\right)\frac{\partial^3 u_0}{\partial x^3} = 0. \tag{3.9b}$$

The above equations define $u_0$, $p_0$ with respect to the time $t_2$ and represent the necessary conditions for obtaining the asymptotic solution of the Boltzmann equation (0.01) with the error $O(\varepsilon)$, which would be valid up to times of $O(\varepsilon^{-2})$. As a result, the terms of order $\varepsilon$ remain bounded when $t$ is as large as $O(\varepsilon^{-2})$, and Eqs. (3.8) become

$$\frac{\partial u_1}{\partial t_1} + \left(\frac{2}{3\lambda_{02}} + \frac{1}{3\lambda_{11}}\right)\frac{\partial^2 \tilde{u}_1}{\partial x^2} = 0, \quad \frac{\partial p_1}{\partial t_1} + \left(\frac{2}{3\lambda_{02}} + \frac{1}{3\lambda_{11}}\right)\frac{\partial^2 \tilde{p}_1}{\partial x^2} = 0. \tag{3.10}$$

It is easy to see that, using the operator (0.15), in which the derivatives $\partial u_0/\partial t_0$, $\partial u_0/\partial t_1$ and $\partial u_0/\partial t_2$ are expressed through the spatial derivatives of (1.2), (2.6) and (3.9), we obtain equations governing hydrodynamic quantities with uniform error $O(\varepsilon)$ valid for $t \leqslant O(\varepsilon^{-2})$

$$\frac{\partial u}{\partial t} + \frac{\partial p}{\partial x} + \frac{\varepsilon}{\lambda_{02}}\left(\frac{2}{3} + \frac{1}{3}\frac{\lambda_{02}}{\lambda_{11}}\right)\frac{\partial^2 u}{\partial x^2} - \frac{\varepsilon^2}{\lambda_{02}^2}\left(\frac{8}{15} - \frac{2\lambda_{02}}{5\lambda_{11}} + \frac{\lambda_{02}^2}{10\lambda_{11}^2}\right)\frac{\partial^3 p}{\partial x^3} = 0,$$

$$\frac{\partial p}{\partial t} + \frac{5}{3}\frac{\partial u}{\partial x} + \frac{\varepsilon}{\lambda_{02}}\left(\frac{2}{3} + \frac{1}{3}\frac{\lambda_{02}}{\lambda_{11}}\right)\frac{\partial^2 p}{\partial x^2} - \frac{\varepsilon^2}{\lambda_{02}^2}\left(\frac{8}{9} - \frac{2\lambda_{02}}{3\lambda_{11}} + \frac{\lambda_{02}^2}{6\lambda_{11}^2}\right)\frac{\partial^3 u}{\partial x^3} = 0, \tag{3.11}$$

$$\frac{\partial s}{\partial t} + \frac{\varepsilon}{\lambda_{11}}\varepsilon\frac{\partial^2 s}{\partial x^2} = 0, \quad u = u_0 + O(\varepsilon), \quad p = p_0 + O(\varepsilon), \quad s = s_0 + O(\varepsilon), \quad t \leqslant O(\varepsilon^{-2}).$$

Since in the Maxwell gas $2\lambda_{02} = 3\lambda_{11}$, then the set (3.11) takes the form as in [16-18, 19].

$$\frac{\partial s}{\partial t} - \frac{3}{2}\mu\frac{\partial^2 s}{\partial x^2} = 0,$$

$$\frac{\partial u}{\partial t} + \frac{\partial p}{\partial x} - \frac{7}{6}\mu\varepsilon\frac{\partial^2 u}{\partial x^2} - \frac{19}{120}\mu^2\varepsilon^2\frac{\partial^3 p}{\partial x^3} = 0, \quad \mu = \frac{1}{|\lambda_{02}|}, \quad \lambda_{02} < 0, \tag{3.12}$$

$$\frac{\partial p}{\partial t} + \frac{5}{3}\frac{\partial u}{\partial x} - \frac{7}{6}\mu\varepsilon\frac{\partial^2 p}{\partial x^2} - \frac{19}{72}\mu^2\varepsilon^2\frac{\partial^3 u}{\partial x^3} = 0,$$

$$u = u_0 + O(\varepsilon), \quad u = p_0 + O(\varepsilon), \quad s = s_0 + O(\varepsilon), \quad t \leqslant O(\varepsilon^{-2}).$$

It should be remarked again that the error in approximations of (3.11) and (3.12) is of $O(\varepsilon)$ when the time interval of interest is not longer than $O(\varepsilon^{-2})$.

If we now introduce new unknown functions $R^+ = a_0 u_0 + p_0$, $R^- = a_0 u_0 - p_0$ (Riemann invariants), then the set (3.11) reduces to the following decoupled equations valid for $t \leqslant O(\varepsilon^{-2})$

$$\frac{\partial s_0}{\partial t} + \frac{\varepsilon}{\lambda_{11}} \frac{\partial^2 s_0}{\partial x^2} = 0,$$

$$\frac{\partial R^+}{\partial t} + a_0 \frac{\partial R^+}{\partial x} + \varepsilon \left( \frac{2}{3 \lambda_{02}} + \frac{1}{3 \lambda_{11}} \right) \frac{\partial^2 R^+}{\partial x^2} - a_0 \varepsilon^2 \left( \frac{8}{15 \lambda_{02}^2} - \frac{2}{5 \lambda_{02} \lambda_{11}} + \frac{1}{10 \lambda_{11}^2} \right) \frac{\partial^3 R^+}{\partial x^3} = 0, \quad (3.13)$$

$$\frac{\partial R^-}{\partial t} - a_0 \frac{\partial R^-}{\partial x} + \varepsilon \left( \frac{2}{3 \lambda_{02}} + \frac{1}{3 \lambda_{11}} \right) \frac{\partial^2 R^-}{\partial x^2} + a_0 \varepsilon^2 \left( \frac{8}{15 \lambda_{02}^2} - \frac{2}{5 \lambda_{02} \lambda_{11}} + \frac{1}{10 \lambda_{11}^2} \right) \frac{\partial R^-}{\partial x^3} = 0.$$

The first corresponds to the heat conduction mode, while the two others of linear Bürgers-Korteweg-de Vries type describe, at the Burnett level, two separate sound waves propagating in either the positive or negative $x$ direction.

For the plane traveling wave in the form $R \exp(\sigma t - ikx)$ the dispersion relations

$$\frac{\sigma}{k} = \frac{\varepsilon k}{\lambda_{11}},$$

$$\frac{\sigma}{k} = \pm i a_0 \left[ 1 + k^2 \varepsilon^2 \left( \frac{8}{15 \lambda_{02}^2} - \frac{2}{5 \lambda_{02} \lambda_{11}} + \frac{1}{10 \lambda_{11}^2} \right) \right] + k \varepsilon \left( \frac{2}{3 \lambda_{02}} + \frac{1}{3} \frac{1}{\lambda_{11}} \right)$$

(3.14)

are identical to the dispersion relations for a plane wave obtained by L. Sirovich from the 13 moments equations by expanding the perturbed distribution function in the eigenfunctions of the collision operator [25].

## Conclusion

D. Hilbert was the first who tried a formal expansion in powers of parameter and extracted a description of the gas in terms of macroscopic quantities. However, as it is known, the direct Hilbert expansion becomes nonuniform as the time tends to infinity and fails, hence, to describe correctly the dissipative processes in the hydrodynamic region. The considered above classical problem shows that multiscale technique allows to overcome these difficulties and to extend the application domain of the Hilbert expansion to the entire interval of dissipative relaxation. At least for the liearized Boltzmann equation. The conditions for eliminating the secular producing terms lead not only to the regular gas-dynamical type equations in lowest approximation and determine the limits of its application, but also to higher-order corrections in the relevant domains.

---------------------------------------------------
\* chivorotnok@web.de
\*\* ichekmarev@web.de
Aachen, Germany